\documentclass{JHEP3} 

\JHEPspecialurl{http://jhep.sissa.it/JOURNAL/JHEP3.tar.gz}
\usepackage{epsfig,multicol,bbm}

\newcommand\fverb{\setbox\fverbbox=\hbox\bgroup\verb}
\newcommand\fverbdo{\egroup\medskip\noindent%
			\fbox{\unhbox\fverbbox}\ }
\newcommand\fverbit{\egroup\item[\fbox{\unhbox\fverbbox}]}
\newbox\fverbbox

\def\vev#1{\left\langle #1\right\rangle}

\def\hbar{\hspace{0pt}\raisebox{1pt}{$-$} \hspace{-7pt} h}

\def\5{\overline 5}

\newcommand{\nn}{\nonumber}

\date{\today}

\preprint{IFIC/07-36}

\author{F.~Bazzocchi, S.~Kaneko and S.~Morisi\\
           E-mail:  \email{fbazzo@ific.uv.es, satoru@ific.uv.es, morisi@ific.uv.es}\\ 
               Instituto de F\'{\i}sica Corpuscular --
  C.S.I.C./Universitat de Val{\`e}ncia \\
  Campus de Paterna, Apt 22085,
  E--46071 Val{\`e}ncia, Spain}

\date{\today}
\title{
A SUSY  $A_4$ model for fermion masses and mixings
}

\abstract{
\noindent  
We study a supersymmetric extension of the Standard Model based on discrete $A_4\times Z_3 \times Z_4$
flavor symmetry. We obtain quark mixing angles as well as a realistic fermion mass spectrum 
and we predict tribimaximal leptonic mixing by a spontaneous breaking of $A_4$. 
The top quark Yukawa interaction is present at the renormalizable level in the superpotential while 
all the other Yukawa interactions arise only at  higher orders. 
We study the Higgs potential and show
that it can potentially solve the so called vacuum alignment problem.
The leading order predictions are not spoiled by subleading corrections.}
\maketitle
\begin{document}


\section{Introduction}
Apart from dark matter and dark energy, the experimental observation of neutrino oscillations is the only evidences of physics beyond 
the Standard Model (SM).
A  global fit of all neutrino data at 2$\sigma$  gives  the following allowed ranges for the  lepton mixing angles\cite{Maltoni:2004ei}:  
\begin{equation}\label{eq:expang}
0.26\le\sin^2\theta_{12}\le 0.36\, ,\,\,0.38\le\sin^2\theta_{23}\le 0.63\, ,\,\,
\sin^2\theta_{13}\le 0.025\,.
\end{equation}
These values are consistent with an especial simple, the so called tribimaximal (TB) mixing ansatz \cite{Harrison:2002er}:
\begin{equation}\label{TB}
\sin^2\theta_{12}=1/3
\, ,\,\,
\sin^2\theta_{23}=1/2
\, ,\,\,
\sin^2\theta_{13}=0
\,.
\end{equation}
As has been shown in \cite{Ma:2001dn,Babu:2002dz,Hirsch:2003dr,Ma:2004zv,Altarelli:2005yp,Chen:2005jm,Zee:2005ut,Altarelli:2005yx,
Adhikary:2006wi,Valle:2006vb,Adhikary:2006jx,Ma:2006vq,Altarelli:2006kg,Hirsch:2007kh,Altarelli:2007cd} this peculiar mixing pattern can be 
explained by an $A_4$ flavor symmetry where $A_4$ is the discrete group of even permutations  
among four objects. The $A_4$ group has four irreducible representations, namely a triplet {\bf 3} and three 
singlets ${\bf 1}$, ${\bf 1'}$ and ${\bf 1''}$ \cite{Luhn:2007uq,Frampton:2000mq}. 
In the majority of  the models based on $A_4$ as  flavor symmetry, the three lepton doublets are assigned to the triplet representation of $A_4$ and the 
three right-handed singlets are assigned to the three $A_4$ singlets.

In spite of the success of the $A_4$ symmetry to explain TB mixing, 
the extension of these models to the quark sector is not straightforward. 
The simplest way to extend the $A_4$ flavor symmetry to the quark sector
is to adopt the same structure as within the lepton sector.
With such an assignment, 
up and down-type quark mass matrices are both diagonalized by the same unitary matrix
giving rise to the diagonal CKM matrix. 
As a first approximation this  leading order result is acceptable.
However, in order to reproduce the correct CKM matrix it is necessary to introduce some subleading terms responsible for inducing correction of order of the Cabibbo angle $\lambda$ in the quark mass matrices and so in the CKM matrix.  
Unluckily it has been shown \cite{Altarelli:2005yx} that the requirement that the subleading terms do not spoil the  leptonic TB mixing matrix forces the corrections in the 
 CKM mixing matrix to be of  $\mathcal{O}(\lambda^2)$ instead of  $\mathcal{O}(\lambda)$. Therefore the corrections coming from subleading terms are too small to explain 
the observed quark mixing angles.
Other possibilities to 
obtain a realistic CKM matrix without spoiling the leptonic TB mixing prediction have been studied in literature. 
An example of $SU(5)$ grand unified case can be found in Ref.\,\cite{Ma:2006sk}. 
Models in which the $A_4$ flavor symmetry both explicitly \cite{Ma:2002yp} and spontaneously broken \cite{He:2006dk}, can also be found in literature.
Another possibility is to consider the discrete groups larger than $A_4$ for flavor symmetry. Recently, some models have been proposed based on the discrete symmetries $T'$ \cite{Carr:2007qw,Feruglio:2007uu} and $\Delta(27)$ \cite{Ma:2006ip,deMedeirosVarzielas:2006fc}.

In this paper, we propose a model based on $A_4$ adopting a different assignment of the representation for the quarks and leptons.
In particular we assign both left and right-handed quarks to the singlets $1,\,1'$ and $1''$ of $A_4$. 
Therefore, we have enough freedom to make the quark sector quite realistic.

As in \cite{Feruglio:2007uu}, in our model only the top quark interaction 
is present at renormalizable level and we also assume the Higgs doublets superfields $\hat{H}_u$ and $\hat{H}_d$ transform as
singlets under $A_4$. 
We use lepton number conservation to make a difference between quark and lepton sectors. 
The charged lepton and quark mass hierarchies are obtained in our model by introducing two auxiliary discrete symmetries $Z_3$ and $Z_4$. 
All the fermion mass hierarchies are realized only by the discrete symmetries. 
Unfortunately the model we propose can not be directly incorporated into a grand unified gauge theory. For recent attempts in this direction see 
\cite{Ma:2006sk,deMedeirosVarzielas:2006fc,Ma:2006wm,King:2006np,Morisi:2007ft,Chen:2007af}. 
The choice of assign both left-handed and right-handed quarks to singlets of $A_4$ is completely different from models previously proposed in 
\cite{Ma:2001dn,Babu:2002dz,Hirsch:2003dr,Altarelli:2005yx,Ma:2002yp,He:2006dk} where left-handed quarks belong to triplet 
representation of $A_4$.
Moreover, most of the $A_4$ and models based on different discrete groups need to invoke additional assumptions to explain fermion mass hierarchies. 
For example a continuous symmetry a la Froggatt-Nielsen.

The read of this paper is organized as follows. In section \ref{sec:model} we describe  the basic features of the model and its field content.
In section \ref{sec:pot} we study the scalar potential and vevs alignment problem. 
In Appendix we consider all possible sources of corrections to the leading textures and show the stability of the leading order predictions.

\section{The model}\label{sec:model}
{\small
\begin{table}[t]
\begin{center}
\begin{tabular}{|c||c|c|c|c||c|c|c|c|c|c|c|c|c||c|c|}
\hline
& $\hat{L}_i$ & $\hat{E}_1^c$ & $\hat{E}_2^c$ & $\hat{E}_3^c$ & $\hat{Q}_1$ & $\hat{Q}_2$ & $\hat{Q}_3$ & $\hat{U}_1^c$ & $\hat{U}_2^c$ & $\hat{U}_3^c$ & $\hat{D}_1^c$& $\hat{D}_2^c$ & $\hat{D}_3^c$ & $\hat{H}_u$ & $\hat{H}_d$\\
\hline
$A_4$ & 3& 1 &1$''$ &1$'$ & 1$''$ & 1$'$ & 1 & 1$''$ & 1$'$ & 1&1& 1$''$ &  1$'$ & $1$ & $1$ \\
$Z_3$& $\omega$&$\omega^2$&$\omega$&$\omega^2$&$\omega$&$\omega$&1&$\omega$&$\omega$&$1$&$\omega^2$&$\omega$&$\omega$ & $1$ & $1$\\
$Z_4$&1&$i$&$-i$&1&1&1&1&1&1&1&1&1&1&1&1\\
$U_L(1)$&1&-1&-1&-1&0&0&0&0&0&0&0&0&0&0&0\\
\hline
\end{tabular}
\caption{The representations of the MSSM matter and Higgs fields under the symmetries associated to $A_4$, $Z_3$ ($\omega^3=1$), $Z_4$ and lepton number $L$. 
The $\hat{L}_i$ and $\hat{Q}_i$ denote the lepton and quark electroweak doublet superfields for $i$-th generation, the $\hat{E}_i^c$, $\hat{U}_i^c$ 
and $\hat{D}_i^c$ denote the electroweak singlet superfields for $i$-th generation, respectively.
The $\hat{H}_u$ and $\hat{H}_d$ are Higgs electroweak doublet superfields. 
}
\label{tab:tab0}
\end{center}
\end{table}
}
{\small
\begin{table}[t]
\label{tab:scal}
\begin{center}
\begin{tabular}{|c||c|c|c|c|c|c|c|c|}
\hline
&$\hat{\phi}_T$ & $\hat{\phi}_S$ &$\hat{\xi}_C,\hat{\xi'}_C$ & $\hat{\xi}$,$\hat{\xi'}$ &$\hat{\chi}$ & $\hat{\theta}$ \\
\hline
$A_4$  &3& 3&1&1 &1$'$ &1$''$\\
$Z_3$&1&$\omega$&$\omega$&$\omega$&$\omega^2$&$\omega^2$\\
$Z_4$&1&1&$i$&1&1&1\\
$U_L(1)$&0&-2&0&-2&0&0\\
\hline
\end{tabular}
\caption{The representations of the additional gauge singlet fields under the symmetries associated to $A_4$, $Z_3$, $Z_4$ and lepton number $L$. 
The $\hat{\phi}_T$, $\hat{\phi}_S$ are  $A_4$ triplets, $\hat{\xi}_C,\hat{\xi'}_C$, $\hat{\xi}$,$\hat{\xi}'$, $\hat{\chi}$ and $\hat{\theta}$ are $A_4$ singlet fields.
}
\label{tab:tab1}
\end{center}
\end{table}
}
We assign the MSSM matter fields and the additional gauge singlet fields to the group representations of 
$A_4\times Z_3 \times Z_4 \times U_L(1)$ respectively 
as shown in Tables \ref{tab:tab0} and \ref{tab:tab1}. 
$\hat{L}_i$ and $\hat{Q}_i$ are the lepton and quark electroweak doublet superfields for $i$-th generation, the $\hat{E}_i^c$, $\hat{U}_i^c$ and $\hat{D}_i^c$ 
are the lepton and quark electroweak singlet superfields for $i$-th generation,
the $\hat{H}_u$ and $\hat{H}_d$ are  the  up and down type Higgs electroweak doublet superfields, respectively.
We also introduce new fields, responsible for flavor symmetry breaking, 
namely $\hat{\phi}_S$ and $\hat{\phi}_T$ as triplets, and 
$\hat{\xi}_C,\hat{\xi'}_C$, $\hat{\xi}$,$\hat{\xi}'$, $\hat{\chi}$ and $\hat{\theta}$ as singlets of $A_4$, respectively.  
In the following we will refer to  the scalars that transform non trivially under the flavor symmetry $A_4\times Z_3 \times Z_4$ 
as \emph{flavons} and analogously we will address the supermultiplets they belong to as  \emph{flavon supermultiplets}.

Once we have assumed that $\hat{H}_u$ and $\hat{H}_d$ transform as singlets under $A_4$, we are forced to put $\hat{Q}_3$ and $\hat{U}^c_3$ 
in the singlet representation {\bf 1} of $A_4$ in order to have a renormalizable interaction for the top quark.
The $\hat{Q}_3-\hat{D}^c_3$ interaction, as well as all the others Yukawa interactions for the up and down type quarks, 
arise from higher dimensional operators. 
This is because we have chosen the charge assignment for the down sector different from the up one:
$\hat{Q}_3$ and $\hat{D}^c_3$ transform as ${\bf 1}$ and ${\bf 1'}$ respectively under $A_4$. 
In particular, we will show in the following 
that quark mixing angles and mass hierarchies are a consequence of the product of the auxiliary symmetry $Z_3$
and the $Z_3$ contained in $A_4$. The  auxiliary $Z_4$ symmetry is introduced to explain charged lepton mass 
hierarchies. Lepton number conservation is crucial to distinguish quark and lepton sectors, namely  
$\hat{\phi}_S$ and $\hat{\xi}$ fields interact with leptons and  not with quarks. 
The field ${\xi'}_C$ is identical to ${\xi}_C$ and 
the reason for its introduction will be come clear in Sec.~\ref{sec:pot}, 
where we will discuss the problem of how to guarantee the correct vacuum alignment.  
Since we can always rotate  ${\xi}_C, {\xi'}_C$  and go to the basis in which just one of the two fields  interacts with the fermions,  
we will neglect the  terms involving  ${\xi'}_C$  in all the Yukawa interactions of the next sections.

\subsection{Leptons}\label{sec:lep}
In the lepton sector, 
there are no Yukawa interactions in the renormalizable superpotential. 
All interactions arise when subleading contributions are considered:
\begin{eqnarray}\label{eq:Ll}
W_{l}&=&
y_e \hat{E}_1^c\,(\hat{\phi}_T \,\hat{L}) \hat{H}_d\,\frac{\hat{\xi}_C\hat{\xi}_C\hat{\xi}_C}{\Lambda^4}+
y_\mu \hat{E}_2^c\,(\hat{\phi_T} \hat{L})' \hat{H}_d\,\frac{\hat{\xi}_C}{\Lambda^2}+
y_\tau \hat{E}_3^c\,(\hat{\phi}_T \,\hat{L})'' \hat{H}_d\frac{1}{\Lambda}\nonumber\\
&+&y_{\nu \xi}\frac{(\hat{L}  \hat{L})\, \hat{H}_u\hat{H}_u\hat{\xi}}{\Lambda^2}+
y_{\nu S}\frac{(\hat{\phi}_S\,\hat{L}\hat{L})\hat{H}_u \hat{H}_u}{\Lambda^2}+h.c. \,.
\end{eqnarray}
Here $\Lambda$ is the cut-off scale, $y_i$ are $\mathcal{O}(1)$ coupling constants
and $(LL)'$, for instance, stands for the product of two triplets of $A_4$ that transforms like a ${\bf 1'}$ singlet (see Appendix \,\ref{conv}). 
In Eq.~\ref{eq:Ll} we have neglected subleading terms of the form
$
\hat{E}_3^c(\hat{\phi}_T \hat{L})''\hat{H}_d\hat{\chi}\hat{\chi}\hat{\chi}/\Lambda^4 
$,
 $
\hat{E}_3^c(\hat{\phi}_T\hat{L})'' \hat{H}_d\hat{\theta}\hat{\theta}\hat{\theta} /\Lambda^4 
$.
The charged lepton mass matrix is given by
\begin{equation}\label{eq:ml}
M^l=U\cdot\left(
\begin{array}{ccc}
y_e u_{\xi_C}^3/\Lambda^3&0&0\\
0& y_\mu u_{\xi_C}/\Lambda&0\\
0& 0&y_\tau 
\end{array}
\right)\frac{v_T v_d}{\Lambda};\ \ 
U=\frac{1}{\sqrt{3}}
\left(
\begin{array}{ccc}
1&1&1\\
1&\omega^2&\omega\\
1&\omega&\omega^2
\end{array}
\right)\,,
\end{equation}
where $u_{\xi_C}$, $v_{T}$ and $v_{d}$ are the vevs in the scalar component of $\hat{\xi}_C$, $\hat{\phi}_T$ and $\hat{H}_{d}$, respectively.
For the charged lepton mass spectrum,  we obtain
\begin{eqnarray}
\label{eq:chldiag}
\frac{m_\mu}{m_\tau}\simeq\frac{u_{\xi_C}}{\Lambda};&\quad &\frac{m_e}{m_\tau}\simeq\left(\frac{u_{\xi_C}}{\Lambda}\right)^3\,,
\end{eqnarray}
and  a realistic hierarchy among the charged lepton masses is recovered by assuming
\begin{equation}
\label{eq:vevxic}
\frac{u_{\xi_C}}{\Lambda} \sim \lambda^2\,,
\end{equation}
with $\lambda$ the Cabibbo angle. 
For the absolute scale of the mass matrix of Eq.~\ref{eq:ml}, assuming $y_\tau\sim 1$ and $v_d\sim 100$ GeV,  one finds
\begin{equation}
\label{eq:vevT}
\lambda^3 \lesssim \frac{v_T}{\Lambda} \lesssim \lambda^2\,.
\end{equation}
From Eq.~\ref{eq:Ll} we obtain the effective neutrino mass matrix as
\begin{equation}\label{eq:mnu}
M^\nu_{LL}=\left(
\begin{array}{ccc}
a&0&0\\
0&a&b\\
0&b&a
\end{array}
\right)
\frac{v_u^2}{\Lambda};\ \ \ 
a=y_{\nu \xi}\,\frac{u_\xi}{\Lambda}
\,,\,\,
b=y_{\nu S}\, \frac{v_S}{\Lambda}\,,
\end{equation}
where $u_{\xi}$, $v_{S}$ and $v_u$ are the vevs of the scalar component of $\hat{\xi}$, $\hat{\phi}_S$ and $\hat{H}_u$, respectively.
In the basis where the charged lepton mass matrix $M_l$ in Eq.~\ref{eq:ml} is diagonal, namely $\hat{L}\,\to \,U \hat{L}$, 
the neutrino mass matrix is diagonalized by the tribimaximal unitary matrix. 
The new ingredient here is that the charged lepton mass hierarchies is derived  from a $Z_3\times Z_4$ discrete symmetry 
and not a Froggatt-Nielsen continuous one.

The scale $\Lambda$ of the model can be fixed by requiring that the neutrino Yukawa couplings are of order $\mathcal{O}(1)$. 
With the plausible assumption\footnote{This assumption will be justified in the discussion of the potential of the model, see next section.} 
of having  $v_S\sim u_\xi \sim v_T $ Eq.~\ref{eq:mnu} then leads
\begin{equation}
 \,\frac{v_u^2 v_S}{\Lambda^2}\simeq 1\,\mbox{eV}\,,
\end{equation}
{\it i.e.} $\Lambda\simeq 10^{10}-10^{12}$ GeV.

\subsection{Quarks}\label{sec:quark}
From the assignments in Tables \ref{tab:tab0} and \ref{tab:tab1}, it follows that the only renormalizable interaction 
in the superpotential is given by 
\begin{eqnarray}\label{eq:Lq0}
W_q^{(0)} &=&y_t\, \hat{Q}_3 \hat{U}^c_3 \hat{H}_u\,,
\end{eqnarray}
giving mass to the top quark. Therefore at leading order only the top quark is massive and all the other quarks are
massless. We will show in this section that the correct masses and mixing pattern is achieved by considering  
subleading corrections to the leading superpotential $W_q^{(0)}$ for quarks in Eq.~\ref{eq:Lq0}. In the next section we will show
that such a result is stable, {\it i.e.} the mass textures do not change once we  consider subleading 
corrections to the superpotential:
\begin{eqnarray}\label{eq:Lq}
W_q 
&=&W_q^{(0)}+(c_u\,\hat{Q}_2 \hat{U}_3^c \hat{\theta} \hat{H}_u+
c'_u\,\hat{Q}_3 \hat{U}_2^c \hat{\theta} \hat{H}_u+ 
b_u\, \hat{Q}_1 \hat{U}_3^c \hat{\chi} \hat{H}_u+
b'_u\, \hat{Q}_3 \hat{U}_1^c \hat{\chi} \hat{H}_u
)/\Lambda\nn\\
&+&
(y_c\,\hat{Q}_2 \hat{U}_2^c \hat{\theta}\hat{\theta} \hat{H}_u+
 a_u\,\hat{Q}_1 \hat{U}_2^c \hat{\theta}\hat{\chi} \hat{H}_u +
a'_u\,\hat{Q}_2 \hat{U}_1^c \hat{\theta}\hat{\chi} \hat{H}_u 
)/\Lambda^2\nn\\
&+&
y_u\,\hat{Q}_1 \hat{U}_1^c \hat{\chi}\hat{\chi} \hat{H}_u
/\Lambda^2\nn\\
&+&
(y_b\,\hat{Q}_3 \hat{D}_3^c \hat{\theta} \hat{H}_d+c_d'\,\hat{Q}_3 \hat{D}_2^c \hat{\chi} \hat{H}_d)/\Lambda\nn\\
&+&
(y_s\,\hat{Q}_2 \hat{D}_2^c \hat{\theta}\hat{\chi} \hat{H}_d+
b_d'\,\hat{Q}_3 \hat{D}_1^c \hat{\theta}\hat{\chi} \hat{H}_d+
b_d\,\hat{Q}_1 \hat{D}_3^c \hat{\theta}\hat{\chi} \hat{H}_d+
c_d\,\hat{Q}_2 \hat{D}_3^c \hat{\theta}\hat{\theta} \hat{H}_d+
a_d\,\hat{Q}_1 \hat{D}_2^c \hat{\chi}\hat{\chi} \hat{H}_d+
)/\Lambda^2\nn\\
&+&(
a'_d\,\hat{Q}_2 \hat{D}_1^c \hat{\chi}\hat{\theta}\hat{\theta} \hat{H}_d+
y_d\,\hat{Q}_1 \hat{D}_1^c \hat{\chi}\hat{\chi}\hat{\theta} \hat{H}_d
)/\Lambda^3\,.
\end{eqnarray}
Here $y_i$, $a^{(')}_i$, $b^{(')}_i$ and $c^{(')}_i$ are coefficients of order $\mathcal{O}(1)$ 
and the following quark mass matrices result
\begin{equation}\label{eq:mu}
M^u=v_u\,\left(
\begin{array}{ccc}
\frac{u_{\chi}^2}{\Lambda^2}y_u & 
\frac{u_\theta u_\chi}{\Lambda^2}a_u&
\frac{u_\chi}{\Lambda}b_u\\
\frac{u_\theta u_\chi}{\Lambda^2} a'_u  & 
\frac{u_\theta^2 }{\Lambda^2} y_c &
\frac{u_\theta}{\Lambda} c_u\\
\frac{u_\chi}{\Lambda} b'_u & 
\frac{u_\theta}{\Lambda} c'_u & y_t
\end{array}
\right),\ \ 
M^{d}=v_d\,
\left(
\begin{array}{ccc}
\frac{u_\theta u_\chi^2}{\Lambda^3}y_d &
\frac{u_\chi^2}{\Lambda^2} a_d & 
\frac{ u_{\chi} u_\theta}{\Lambda^2} b_d \\
\frac{u_\theta^2 u_\chi}{\Lambda^3}a'_d & 
\frac{u_\theta u_\chi }{\Lambda^2} y_s & 
\frac{u_\theta^2 }{\Lambda^2} c_d \\ 
\frac{u_\theta u_\chi}{\Lambda^2}b'_d & 
\frac{u_\chi}{\Lambda} c'_d & 
\frac{u_\theta}{\Lambda}y_b 
\end{array}
\right)\,.
\end{equation}
Here $u_{\theta}$ and $u_{\chi}$ are the vevs of the scalar component of $\hat{\theta}$ and $\hat{\chi}$.
In Eqs.~\ref{eq:mu} the up and  down quark mass matrices are correctly reproduced for a natural choice of the parameters: 
\begin{eqnarray}
\label{eq:vevq}
 \frac{u_\chi}{\Lambda} \sim  \lambda^3\,,&\quad& 
\frac{u_\theta}{\Lambda}\sim \lambda^2 \,.
\end{eqnarray}
With the choice of Eq.~\ref{eq:vevq},  the quark masses and mixing angles are approximately given by
\begin{eqnarray}\label{eq:evud}
m_u\approx y_u\frac{u_{\chi}^2  }{\Lambda^2}v_u \sim \lambda^6 v_u\,,\quad& 
m_c\approx  y_c\frac{u_\theta^2}{\Lambda^2} v_u \sim \lambda^4v_u\,, \quad&
m_t\approx y_t v_u \sim v_u, \nn\\
m_d \approx  y_d\frac{u_\theta u_\chi^2}{\Lambda^3} v_d \sim \lambda^8v_d\,, \quad& 
m_s\approx y_s \frac{u_\theta u_\chi }{\Lambda^2}  v_d \sim \lambda^5v_d\,, \quad&
m_b\approx \frac{u_\theta}{\Lambda} y_b v_d \sim \lambda^2v_d\,,
\end{eqnarray}
and the CKM matrix elements can be estimated as 
\begin{eqnarray}\label{eq:VuVd}
V_{us}\approx \frac{u_\chi}{u_\theta}\left(\frac{a_d}{y_s}-\frac{a_u}{y_c} \right)\sim \lambda;  \quad
V_{ub}\approx \frac{u_\chi}{\Lambda}\left( \frac{b_d}{y_b}-\frac{b_u}{y_t}\right)\sim \lambda^3; \quad
V_{cb}\approx \frac{u_\theta}{\Lambda} \left(\frac{c_d}{y_b}- \frac{c_u}{y_t}\right)\sim \lambda^2\,.\nonumber\\
\end{eqnarray}
The quark mass hierarchies of eq.(\ref{eq:evud}) are in slight in contrast with the experimental data. However the free parameters of $\mathcal{O}(1)$
in the quark mass matrices in eq.(\ref{eq:mu}), allow us to fit all the quark masses by admitting a fine tuning of $10\% $ and a reasonable values of $\tan\beta$.


\section{The potential}
\label{sec:pot} 
In Sec.~\ref{sec:lep} we have seen that the TB mixing matrix  is obtained   when the triplets $\hat{\phi}_T$ and $\hat{\phi}_S$ 
break  $A_4$ in the direction $(v_T,v_T,v_T)$ and $(v_S,0,0)$  respectively. 
Our potential must break $A_4$ in this direction.

In Sec.~\ref{sec:model} we have  implicitly assumed that the part of the superpotential  that gives rise to the  Yukawa Lagrangian of the  
SM matter fields is given by  Eq.~\ref{eq:Ll}. This part was obtained by integrating out some heavy fields. 
Inspection of  Eq.~\ref{eq:Ll} show that  $W_l$  is charged under a global $U(1)_R$ symmetry with $R$-charge $+2$ if we assign to 
the matter supermultiplets $R$-charge $+1$ and to the flavon supermultiplets  involved in the effective superpotential $W_l$  $R$-charge $0$.  
The global $U(1)_R$ is the continuous symmetry  that is broken to   the discrete $R$-parity once we include the gaugino masses into the model. 
We may assume that  $W_l$ and the part of the superpotential that involves all the supermultiplets  that transform non trivially under the 
flavor symmetry $A_4\times Z_3 \times Z_4$ has $R$-charge $+2$ and therefore give rise to a Yukawa  superpotential as in Eq.~\ref{eq:Ll} invariant  
with respect to the continuous $U(1)_R$. 
Since the flavon supermultiplets  bring null  $R$-charge, to build a superpotential with total $R$-charge $+2$  avoiding the spontaneous 
breaking of the $R$-symmetry, we need some messenger fields that carry $R$-charge $+2$ is necessary that the superpotential be linear   
in these fields. To the  flavon supermultiplets of Table~\ref{tab:tab1}  we therefore add $4$ messenger  $R$-supermultiplets given in Table~\ref{tab:messhy} .
{\small
\begin{table}[t]
\label{tab:messhy} 
\begin{center}
\begin{tabular}{|c||c|c|c|c|}
\hline
& $\hat{\phi}^T_R$ & $\hat{\phi}^S_R$& $\hat{\phi}^{'S}_R$  & $\hat{\xi}_{R}^C$  \\
\hline
$A_4$ &$3$&$ 3$&$3$&$1$\\
$Z_3$&$1$& $1$&$\omega$&$\omega^2$\\
$Z_4$&$1$&$1$&$1$&$-i$\\
$U_L(1)$&$0$&$2$&$4$&$0$\\
$U_R(1)$&$2$&$2$&$2$&$2$\\
\hline
\end{tabular}
\begin{tabular}{|c|c|c|c|c|c|c|c|}
\hline
$\hat{\phi}_T$ & $\hat{\phi}_S$ &$\hat{\xi}_C,\hat{\xi'}_C$ & $\hat{\xi}$,$\hat{\xi'}$ &$\hat{\chi}$ & $\hat{\theta}$ \\
\hline
3& 3&1&1 &1$'$ &1$''$\\
1&$\omega$&$\omega$&$\omega$&$\omega^2$&$\omega^2$\\
1&1&$i$&1&1&1\\
0&-2&0&-2&0&0\\
0&0&0&0&0&0\\
\hline
\end{tabular}
\caption{The representations of messenger R-supermultiplets and the flavon supermultiplets. }
\end{center}
\end{table}
}
The superpotential $W$ is invariant under all the symmetries  of the model 
that contain an explicit breaking term of the continuous   $U_L(1)$ and the discrete $Z_3$ symmetries.  
We assume that the explicit breaking term arises from the presence of an heavy sector that does not interact with the 
matter content of the model, thus ensuring that all the   terms  generated  in the Yukawa superpotential 
by the $U_L(1)$  and $Z_3$ explicit breaking terms are sufficiently suppressed and do not affect the  fermion mass matrix structures. 
The spontaneous breaking of the continuous lepton number $U(1)_L$ gives to unphysical massless Goldstone boson. 
We will discuss in detail in the Appendix the problem of the Goldstone boson.
Nevertheless, as shown in the Appendix,
by imposing the $R$-symmetry invariance under the discrete symmetries, our superpotential 
presents some accidental global continuous  symmetries that give massless states.  
We guarantee the stability of the minimum and give positive mass to the massless states by adequately choosing the $V_{soft}$ of SUSY.
The full superpotential $W$ is given by
\begin{eqnarray}
\label{eq:superP}
W&=& M_T \,\hat{\phi}_R^T \hat{\phi}_T +\lambda_T \,\hat{\phi}_R^T \hat{\phi}_T  \hat{\phi}_T\nn\\
&+& \lambda_{S\theta} \,( \hat{\phi}_R^S \hat{\phi}_S)'\, \hat{\theta} + \lambda_{S\chi} \,( \hat{\phi}_R^S \hat{\phi}_S)''\,\hat{\chi}+M_C \,\hat{\xi}_R^C \hat{\xi}_C+{M'}_{C}\,\hat{\xi}_R^C \hat{\xi'}_C  \nn \\
&+& 
\lambda_S \,\hat{\phi}_R^{'S}   \hat{\phi}_S \hat{\phi}_S
+\lambda_{\xi} \,\hat{\phi}_R^{'S}   \hat{\phi}_S \hat{\xi}
+\lambda_{\xi}' \,\hat{\phi}_R^{'S}   \hat{\phi}_S \hat{\xi}'\,.
\end{eqnarray}
From Eq.~\ref{eq:superP} we derive  the scalar potential by
\begin{equation}
\label{eq:fullpot}
V=\left |\frac{\partial W}{\partial f_{\phi_i}}\right |^2+ V_{soft}\,,
\end{equation}
where $V_{soft}$ includes all possible SUSY soft terms for the new scalars of the model  invariant under all the discrete($A_4,Z_3,Z_4$) 
and  the continuous ($U_L(1)$) symmetries. It breaks the accidental global continuous symmetries presented by the SUSY invariant scalar potential.  
We will discuss this point in detail in the  Appendix.   Since the scale of $V_{soft}$ is $1-10$ TeV, while the scale of the SUSY 
invariant potential is  $10^{10}-10^{12}$ GeV,   we can neglect  $V_{soft}$ and search for a  vacuum configuration that is SUSY invariant. It is given by
\begin{eqnarray}
\label{eq:susycons}
\vev{\phi_T}&=&(v_T,v_T,v_T)\,,\nn\\
\vev{\phi_S}&=&(v_S,0,0)\,,\nn\\
\vev{\chi}&=& u_\chi \,,\nn\\
\vev{\xi}&=& u_\xi \,, \nn\\
\vev{\xi'}&=& u_\xi' \,, \nn\\
\vev{\xi_C}&=& u_{\xi_C} \,,\nn\\
\vev{{\xi'}_C}&=& u_{{\xi'}_C} \,,\nn\\
\vev{\theta}&=& u_\theta \,. 
\end{eqnarray}
Next we want to identify the region of the parameter space for which the vacuum configuration of Eq.~\ref{eq:susycons} is the minimum of 
Eq.~\ref{eq:fullpot} with $V_{soft}=0$. 
Since none of the messenger fields acquires a vev and since they enter linearly in all the terms of 
the superpotential, in the SUSY limit all the derivatives with respect to the $F$ components of the supermultiplets not charged with 
respect to the $U(1)_R$ symmetry vanish. Therefore in the discussion of the vacuum configuration we have to take into account only  the 
derivatives with respect to the $F$ components of the  messenger supermultiplets. Taking the derivative of $W$ with respect to the 
$F$ components of the supermultiplets  $\hat{\phi}^T_R$, $\hat{\phi}^S_R$ and $\hat{\phi}^{'S}_R$ and substituting the vacuum 
configuration of Eq.~\ref{eq:susycons} we find
\begin{eqnarray}
\label{eq:derW1}
\frac{\partial W}{\partial f_{{\phi^T_R}_i}} &=&\frac{M_T}{3} v_T+ \frac{1}{3} \lambda_{T} v_T^2 \,,\nn\\
\frac{\partial W}{\partial f_{{\phi^S_R}_1}} &=&\frac{\lambda_{S\theta}}{3}v_S u_\theta +\frac{\lambda_{S\chi}}{3}v_S u_\chi \,,\nn\\
 \frac{\partial W}{\partial f_{{\phi^S_R}_{2,3}}} &=&0 \,,\nn\\
\frac{\partial W}{\partial f_{{\phi^{'S}_R}_1}} &=&\lambda_{\xi}  v_S  u_\xi +\lambda_{\xi}'  v_S  u_\xi' \nn\\
\frac{\partial W}{\partial f_{{\phi^{'S}_R}_{1,2}}} &=&0 \,.
 \end{eqnarray}
From Eq.~\ref{eq:derW1} one finds that a possible solution conserving SUSY is given by
\begin{eqnarray}
\label{eq:vev1}
v_T&=&- \frac{M_T}{\lambda_T} \,,\nn\\
u_\chi&=& - \frac{\lambda_{S \theta} }{\lambda_{S \chi} } u_\theta\,, \nn\\
u_\xi&=&-\frac{\lambda_\xi'}{\lambda_\xi}u_\xi' \,.
\end{eqnarray}
Taking the derivative with respect to the $F$ component of $\hat{\xi}^C_R$ we have
\begin{eqnarray}
\label{eq;cond2}
\frac{\partial W}{\partial f_{\xi^C_R}} &=& \lambda_C u_{\xi_C}+ {\lambda'}_{C} u_{{\xi'}_C}\,,
\end{eqnarray}
from which it follows that
\begin{equation}
\label{eq:vev2}
u_{\xi_C}=-\frac{\lambda^{'}_{C}}{\lambda_C} \, u_{{\xi'}_C}\,.
\end{equation}

A few comments might be in order.
In the Secs.~\ref{sec:lep}--\ref{sec:quark}  we have seen that  the correct fermion mass  matrices are obtained assuming that the vevs satisfy 
\begin{eqnarray}
\label{eq:assumvevs}
\frac{u_\theta}{\Lambda}\sim \frac{u_{\xi_C}}{\Lambda} &\sim &\lambda^2,\nn\,\\
\frac{v_T}{\Lambda} \sim \frac{v_S}{\Lambda}\sim \frac{u_\xi}{\Lambda}&\sim&\lambda^2 \div  \lambda^3,\nn\,\\
 \frac{u_\chi}{\Lambda} &\sim&  \lambda^3\,,
\end{eqnarray}
with $\lambda$ the Cabibbo angle. 

From Eqs.~\ref{eq:vev1}  we see that in order to  satisfy the relations given in Eq.~\ref{eq:assumvevs} we need $M_T \sim \lambda^2 \Lambda$, 
$ \frac{\lambda_{S \theta} }{\lambda_{S \chi} } \sim \lambda$.  Aside from $v_T$, all the vevs  remain  undetermined and their natural value 
tends to be $\Lambda$ (the cut-off scale) and not $(\lambda^2 \div \lambda^3)\Lambda$. 
The problem of how to stabilize the relations given in Eq.~\ref{eq:assumvevs} is therefore not completely solved. 
We assume that it can be solved by including loop-contributions or explicit breaking terms  of the abelian discrete symmetry arising in a hidden scalar sector. 
We do not enter into the details of this problem and assume that there exists a choice of the parameters of the potential  
that satisfy the  relations given in Eq.~\ref{eq:assumvevs}.  Once we include $V_{soft}$ we  have to choose the soft terms  in such a way 
that $V_{soft}<0$  in the vacuum configuration  of Eq.~\ref{eq:assumvevs}. The stability of the minimum of the potential of Eq.~\ref{eq:fullpot} can be assured.

Corrections to the leading mass matrix textures of Sec.~\ref{sec:model} are
induced by higher order operators in the superpotential of Eq.~\ref{eq:superP} that 
can change the vacuum configuration given in Eq.~\ref{eq:susycons}. These possibility could disalign the triplet vevs or 
may be directly affect the mass matrices. 
In Appendix we accomplish the full analysis of higher order corrections to the  mass matrix textures.
In particular
we check that the corrections induced by the scalar interactions and by the introduction of 
higher order operators in the superpotential are under control and 
do not destroy the leading order predictions of the model.

\section{Conclusions}
We have  proposed a supersymmetric extension of the standard model based on the discrete flavor group
$A_4$. The new features with respect to earlier work present in the literature is that all the  quarks, both left-handed and right-handed, 
transform as singlet representations of $A_4$,  {\bf 1}, {\bf 1}$'$ and {\bf 1}$''$. 
This assignment  allows us to ensure that only the top quark acquires mass  at tree level.  
All other entries in the mass matrices are generated by higher dimensional operators and are suppressed by powers of $1/\Lambda$. 
The introduction of  two auxiliary discrete symmetries $Z_3$ and $Z_4$ have allowed us to obtain realistic charged fermion  
mass hierarchies  and the CKM  mixing matrix, without appealing to a continuous $U(1)_F$ flavor symmetry, 
lepton number avoids dangerous mixing between quarks and leptons.
Finally, we have studied the scalar superpotential that presents the correct $A_4$ breaking alignments and with the introduction 
of apposite explicit breaking terms of the abelian symmetry of the model we have determined all the flavon scalar vevs.

\acknowledgments

Work supported by MEC grants FPA2005-01269 and FPA2005-25348-E, by Generalitat Valenciana ACOMP06/154, by European Commission Contracts  
MRTN-CT-2004-503369 and ILIAS/N6 WP1 RII3-CT-2004-506222. F.~B. is supported by  MEC postdoctoral grant SB-2005-0161. We thank M.\,Hirsch
for reading the paper.

\bibliographystyle{h-elsevier}
  \bibliography{ref}

\begin{thebibliography}{10}

\bibitem{Maltoni:2004ei}
M. Maltoni et~al.,
\newblock New J. Phys. 6 (2004) 122, hep-ph/0405172.

\bibitem{Harrison:2002er}
P.F. Harrison, D.H. Perkins and W.G. Scott,
\newblock Phys. Lett. B530 (2002) 167, hep-ph/0202074.

\bibitem{Ma:2001dn}
E. Ma and G. Rajasekaran,
\newblock Phys. Rev. D64 (2001) 113012, hep-ph/0106291.

\bibitem{Babu:2002dz}
K.S. Babu, E. Ma and J.W.F. Valle,
\newblock Phys. Lett. B552 (2003) 207, hep-ph/0206292.

\bibitem{Hirsch:2003dr}
M. Hirsch et~al.,
\newblock Phys. Rev. D69 (2004) 093006, hep-ph/0312265.

\bibitem{Ma:2004zv}
E. Ma,
\newblock Phys. Rev. D70 (2004) 031901, hep-ph/0404199.

\bibitem{Altarelli:2005yp}
G. Altarelli and F. Feruglio,
\newblock Nucl. Phys. B720 (2005) 64, hep-ph/0504165.

\bibitem{Chen:2005jm}
S.L. Chen, M. Frigerio and E. Ma,
\newblock Nucl. Phys. B724 (2005) 423, hep-ph/0504181.

\bibitem{Zee:2005ut}
A. Zee,
\newblock Phys. Lett. B630 (2005) 58, hep-ph/0508278.

\bibitem{Altarelli:2005yx}
G. Altarelli and F. Feruglio,
\newblock Nucl. Phys. B741 (2006) 215, hep-ph/0512103.

\bibitem{Adhikary:2006wi}
B. Adhikary et~al.,
\newblock Phys. Lett. B638 (2006) 345, hep-ph/0603059.

\bibitem{Valle:2006vb}
J.W.F. Valle,
\newblock J. Phys. Conf. Ser. 53 (2006) 473, hep-ph/0608101.

\bibitem{Adhikary:2006jx}
B. Adhikary and A. Ghosal,
\newblock Phys. Rev. D75 (2007) 073020, hep-ph/0609193.

\bibitem{Ma:2006vq}
E. Ma,
\newblock Mod. Phys. Lett. A22 (2007) 101, hep-ph/0610342.

\bibitem{Altarelli:2006kg}
G. Altarelli, F. Feruglio and Y. Lin,
\newblock Nucl. Phys. B775 (2007) 31, hep-ph/0610165.

\bibitem{Hirsch:2007kh}
M. Hirsch et~al.,
\newblock (0300), hep-ph/0703046.

\bibitem{Altarelli:2007cd}
G. Altarelli,
\newblock (0500), arXiv:0705.0860 [hep-ph].

\bibitem{Luhn:2007uq}
C. Luhn, S. Nasri and P. Ramond,
\newblock (0100), hep-th/0701188.

\bibitem{Frampton:2000mq}
P.H. Frampton and T.W. Kephart,
\newblock Phys. Rev. D64 (2001) 086007, hep-th/0011186.

\bibitem{Ma:2006sk}
E. Ma, H. Sawanaka and M. Tanimoto,
\newblock Phys. Lett. B641 (2006) 301, hep-ph/0606103.

\bibitem{Ma:2002yp}
E. Ma,
\newblock Mod. Phys. Lett. A17 (2002) 627, hep-ph/0203238.

\bibitem{He:2006dk}
X.G. He, Y.Y. Keum and R.R. Volkas,
\newblock JHEP 04 (2006) 039, hep-ph/0601001.

\bibitem{Carr:2007qw}
P.D. Carr and P.H. Frampton,
\newblock (2007), hep-ph/0701034.

\bibitem{Feruglio:2007uu}
F. Feruglio et~al.,
\newblock (2007), hep-ph/0702194.

\bibitem{Ma:2006ip}
E. Ma,
\newblock Mod. Phys. Lett. A21 (2006) 1917, hep-ph/0607056.

\bibitem{deMedeirosVarzielas:2006fc}
I. de~Medeiros~Varzielas, S.F. King and G.G. Ross,
\newblock Phys. Lett. B648 (2007) 201, hep-ph/0607045.

\bibitem{Ma:2006wm}
E. Ma,
\newblock Mod. Phys. Lett. A21 (2006) 2931, hep-ph/0607190.

\bibitem{King:2006np}
S.F. King and M. Malinsky,
\newblock Phys. Lett. B645 (2007) 351, hep-ph/0610250.

\bibitem{Morisi:2007ft}
S. Morisi, M. Picariello and E. Torrente-Lujan,
\newblock Phys. Rev. D75 (2007) 075015, hep-ph/0702034.

\bibitem{Chen:2007af}
M.C. Chen and K.T. Mahanthappa,
\newblock (0500), arXiv:0705.0714 [hep-ph].

\bibitem{Bazzocchi:2004dw}
F. Bazzocchi,
\newblock Phys. Rev. D70 (2004) 013002, hep-ph/0401105.

\bibitem{Buras:2003jf}
A.J. Buras,
\newblock Acta Phys. Polon. B34 (2003) 5615, hep-ph/0310208.

\end{thebibliography}

\appendix
\section{Conventions}\label{conv}

The finite  group of the even permutations of  four objects is $A_4$ \cite{Luhn:2007uq,Frampton:2000mq}. 
Its generators $S$ and $T$  obey the relations
$$ S^2=(ST)^3=T^3=1.$$
We remark that $A_4$ has four irreducible representations, 3, 1, $1'$ and $1''$ satisfying the following product rule
\begin{eqnarray}
3\times 3&=&3+3+1+1'+1''\nn\\
1\times 1&=&1,\,\,1'\times 1''=1,\,\,1'\times 1'=1'',...\,\,.
\end{eqnarray}
We chose to work in the basis in which $S$ is a diagonal matrix in the three-dimensional representation.
Different basis, like the one where $T$ is diagonal, are related by unitary transformations, see for instance \cite{Altarelli:2007cd}.
Therefore if $a=(a_1,a_2,a_3)$ and $b=(b_1,b_2,b_3)$ are two triplets of $A_4$ their products are given by
\begin{equation} 
\begin{array}{llll}
1&=&(ab)&=\frac{1}{\sqrt{3}}(a_1b_1+ a_2b_2+a_3b_3),\\
1'&=&(ab)'&=\frac{1}{\sqrt{3}}(a_1b_1+ \omega^2 a_2b_2+ \omega a_3b_3),\\
1''&=&(ab)''&=\frac{1}{\sqrt{3}}(a_1b_1+\omega a_2 b_2+\omega^2 a_3b_3),\\
3&=&(ab)_3&=\frac{1}{\sqrt{3}}(a_2b_3,a_3b_1, a_1b_2),\\
3'&=&(ab)'_3&=\frac{1}{\sqrt{3}}(a_3b_2,a_1b_3,a_2 b_1).
\end{array}
\end{equation}

\section{The scalar potential}
\label{sec:potesteso}
The full expression of the scalar  potential at the leading order is given by Eq.\,(\ref{eq:fullpot})
\begin{equation}
V= V_{SUSY}+V_{soft}\,.
\end{equation}
Here $V_{soft}$ contains soft SUSY breaking terms that stabilize the minimum of $V_{SUSY}$. 
From Eqs.~\ref{eq:superP}  the part of the scalar potential that involves  only the flavons is given by 
\begin{eqnarray}
\label{eq:vsusy}
V_{SUSY}&=&| M_T \phi_T + 2 \lambda \phi_T \phi_T |^2+|\lambda_{S\theta}\phi_S\theta+\lambda_{S\chi} \phi_S \chi|^2 \nn\\
&+&|  \lambda_S \phi_S \phi_S + \lambda_\xi' \phi_S \xi'+\lambda_\xi \phi_S \xi |^2+ |M_C \xi_C +{M'}_C {\xi'}_C  |^2 \,.\end{eqnarray}

 Without lost of generality, we can rotate the $\xi$ and $\xi'$ fields in new combinations $\tilde{\xi}'$ and $\tilde{\xi}$
in which only $\tilde{\xi}'$ develop a vev. The same argumentation can be applied
to $\xi_C$ and $\xi'_C$ obtaining   $\tilde{\xi}_C$ and $\tilde{\xi}'_C$.  In terms of the new fields the potential of Eq.~\ref{eq:vsusy} is given by
  \begin{eqnarray}
\label{eq:vsus2y}
V_{SUSY}&=&| M_T \phi_T + 2 \lambda \phi_T \phi_T |^2+|\lambda_{S\theta}\phi_S\theta+\lambda_{S\chi} \phi_S \chi|^2 \nn\\
&+&| \lambda_S \phi_S \phi_S+\tilde{\lambda}_\xi \phi_S \tilde{\xi}|^2+ |\tilde{M}_C \tilde{\xi}_C   |^2 \,.\end{eqnarray}
  with  $\tilde{\lambda}_\xi=\sqrt{\lambda^2_\xi +\lambda^{'2}_\xi }$ and $\tilde{M}_C =\sqrt{M_C^2+M_C^{'2}}$. 
  At this level  the two combination that develop vevs, $\tilde{\xi}'$ and $\tilde{\xi}'_C$,  are massless,  lacking of a potential. They  acquire a (positive) mass only when we add to the  SUSY scalar potential of   Eq.~\ref{eq:vsus2y}  the  soft scalar potential  $V_{soft}$, provided it contains adequate mass terms, as it has been done in eq.\,(\ref{eq:Vsoft}) below. 
The rotation done for $\xi,\xi'$ and $\xi_C,\xi'_C$ is not allowed for $\chi$ and $\theta$ since this two scalar fields  behave differently under $A_4$.  
The terms given by
\begin{equation}
\label{eq:Vs}
| \lambda_S \phi_S \phi_S+\tilde{\lambda}_\xi \phi_S \tilde{\xi}|^2+ |\lambda_{S\theta}\phi_S\theta+\lambda_{S\chi} \phi_S \chi|^2 \,,
\end{equation}
present  two accidental continuous  global  symmetry. 
The first term in Eq.\,(\ref{eq:Vs}) $$| \lambda_S \phi_S \phi_S+\tilde{\lambda}_\xi \phi_S \tilde{\xi}|^2$$ 
presents an accidental continuous symmetry  $SO(3)\times U(1)_L\sim O(3)$.
With respect to $SO(3)$, $\phi_S$ transforms as a triplet, while $\tilde{\xi},\chi$ and $\theta$  as  singlets.
When $\phi_S$ acquires vev as $\vev{\phi_S}=(v_S,0,0)$ the accidental continuous symmetry is   broken to $SO(2) \times U(1)\sim O(2)$ leaving two massless Goldstone bosons.  
Above the two fields charged under the lepton number $U(1)_L$, $\phi_S$ and $\tilde{\xi}$, only $\phi_S$ develops a vev. As consequence, the   Goldstone boson associated to the breaking of  the abelian $U(1)_L$  has projection only along $\phi_S$ as can be seen in Eq.~\ref{eq:ms}.
The real  parts of $\chi$ and $\theta$ belong to a doublet of $O(2)$ and the imaginary components of $\chi$ and $\theta$ belong to another doublet of $O(2)$.  
When $\chi$ and $\theta$ develop vevs, the $O(2)$ global symmetry is broken and other two massless Goldstone bosons arise. 
Indeed  the mass matrices for the real and imaginary components of the scalars involved in the part of the scalar potential of Eq.~\ref{eq:Vs}  are identical. 
In the basis $(\phi_{S_1},\,\phi_{S_2},\,\phi_{S_3},\,\tilde{\xi},\,\chi,\,\theta)$  the mass matrix is given by
\begin{eqnarray}\label{eq:ms}
\mathcal{M}^2_S&=&\left(
\begin{array}{cccccc}
0&0&0&0&0&0\\
0  & \frac{2}{3}( \lambda_S v_S^2 + \lambda_{S\theta} u_\theta^2) &0&0&0 &0  \\
0  &0   & \frac{2}{3}( \lambda_S v_S^2 + \lambda_{S\theta} u_\theta^2) &0&0&0  \\
0  &  0 &   0& \frac{2}{3}(\lambda^2_\xi +\lambda^{'2}_\xi ) v_S^2 & 0&0\\
0&0&0&0& \frac{2}{9} \lambda_{S\chi}^2 v_S^2& \frac{2}{9} \lambda_{S\chi}\lambda_{S\theta} v_S^2\\
0&0&0&0&  \frac{2}{9} \lambda_{S\chi}\lambda_{S\theta} v_S^2&\frac{2}{9} \lambda_{S\theta}^2 v_S^2
\end{array}
\right)\,.\nn\\
&&
 \end{eqnarray}
We recognize $1+1$  Goldstone bosons from the entry (1,1) of the matrix $\mathcal{M}^2_S$.
The determinant of the sub-block 5,6 of the matrix $\mathcal{M}^2_S$ is zero, so $\mathcal{M}^2_S$ gives other $1+1$
Goldstone bosons.

The first term in the potential of Eq.~\ref{eq:vsusy}  is
 \begin{equation}
 \label{eq:potT}
| M_T \phi_T + 2 \lambda \phi_T \phi_T |^2.
\end{equation}
It is invariant under an accidental $SO(3)$ symmetry under which $\phi_T$ transform as a triplet. 
When $\phi_T$ acquire vev along the direction $(1,1,1)$, $SO(3)$ is broken into $S_3$
giving rise to the following mass matrix for the real (imaginary) components of $\phi_{T}$
\begin{eqnarray}
\label{eq:masst}
\mathcal{M}_T^2 &=& 
8 \lambda_T^2 v_T^2 \,\left(
\begin{array}{ccc}
 1 &- \frac{1}{3}  &  -\frac{1}{3} \\
-\frac{1}{3}  & 1  &-  \frac{1}{3} \\
-  \frac{1}{3}&-  \frac{1}{3} &   1
\end{array}
\right)\,.
\end{eqnarray}
Since the mass matrix of  Eq.~\ref{eq:masst}  is $S_3$ invariant the mass eigenstates are one singlet of  $S_3$, with  
mass  given by $\frac{8}{3}\lambda_T^2 v_T^2$,  and one $S_3$ doublet, with  mass  $\frac{32}{3}\lambda_T^2 v_T^2$.
The absence of Goldstone bosons suggests that when $\phi_T$ acquires vev in the direction  $(1,1,1)$  the continuous accidental global 
symmetry $SO(3)$   is broken to  a  continuous group of its same dimensions, 3, and that contains $S_3$. This group is $SU(2)$ and the 
doublet of $S_3$ transforms as a doublet of $SU(2)$.

The scalar potential of Eq.\,(\ref{eq:vsusy}) gives in total $4+4$ massless particles.
We break explicitly  the accidental global symmetries and the continuous lepton number $U(1)_L$ in $V_{soft}$
in order to give masses to these 8 massless particles.
The  breaking terms have to provide positive mass to the unwanted Goldstone bosons and in  general this is not an easy issue  since $V_{soft}$ has to be negative
 in the minimum in order to guarantee the stability of the potential. 
  Indeed the problem related to the presence of  flavor Goldstone bosons  is given by  the  
$4$-fermion effective operator \cite{Bazzocchi:2004dw,Buras:2003jf}. 
This effective operator may be parametrized by
\begin{equation}
\label{eq:4f}
\frac{1}{m_\alpha^2} \frac{m^f_{ij}m^f_{kl}}{v^2_F } \bar{\psi}_i (c^{ij}_V+ \gamma_5 c^{ij}_A)\psi_j \,\bar{\psi}_k (c^{kl}_V+ \gamma_5 c^{kl}_A)\psi_l
\end{equation}
where $m^f_{ij}$ is the fermion mass matrix entry, $v_F$ the flavor symmetry breaking scale and $m_\alpha$ the mass of the scalar or pseudoscalar flavon that mediates the process that gives rise to the  $4$-fermion operator.  If we want to give a very rough and conservative estimate of the mass $m_\alpha$ needed to suppress flavor changing processes we can overestimate the operator of Eq.~\ref{eq:4f}
with
\begin{equation}
\label{eq:4fn}
\frac{1}{m_\alpha^2} \frac{m_3^{f\,2}}{v_F^2 } \bar{\psi}_i \psi_j \,\bar{\psi}_k \psi_l\,,
\end{equation}
where $m_3^f$ is the heaviest mass in the $f$ fermion family.  If we now consider for example the process $\mu \to 3 e$ we obtain that  the operator given in Eq.~\ref{eq:4fn} gives a decay width
\begin{equation}
\label{eq:gmu}
\Gamma_{\mu \to 3 e} \sim \frac{1}{16 \pi^3} \frac{m_\mu^5}{m_\alpha^4} \frac{m_\tau^4}{v_F^4}\,,
\end{equation}
and by comparing Eq.~\ref{eq:gmu} with 
\begin{equation}
\Gamma_{\mu \to all} =\frac{m_\mu^5 G_F^2}{192 \pi^3} \,,
\end{equation}
and with the experimental bound  $\Gamma_{\mu \to 3 e} /\Gamma_{\mu \to all}< 10^{-12}$,  we obtain for a  flavor  breaking scale around  $10^{10}$ GeV the very  low bound
\begin{equation}
m_\alpha \sim 1 \mbox{GeV}\,.
\end{equation}
The same bound is obtained by considering other tree level flavor changing processes like $K-\bar{K}$ oscillation.
If we assume that the SUSY breaking scale of the soft potential $V_{soft}$ is around the TeV, then  the inclusion in $V_{soft}$ of terms that give a positive mass around the GeV to the unwanted Goldstone bosons does not endanger the stability of the potential.   However we lack of a   dynamical principle to justify why the scale of some  soft terms  is around the TeV  while that  of the others is around the GeV and therefore we prefer  assuming that all the soft terms are around $100$ GeV- $1$ TeV and imposing the constrain $V_{soft}<0$ on the parameter of $V_{soft}$.  For example the soft potential  $V_{soft}$

\begin{eqnarray}
\label{eq:Vsoft}
V_{soft}&=& \frac{m_T}{2}^2(\phi_T^2+ H.c.)+\frac{A_T}{3}(\phi_T^3+ H.c.)+\tilde{m}_T^2 |\phi_T|^2 \nn\\     
&+& A_S [  (\phi_S\phi_S)'\chi +H.c.]  +m^2_{\tilde{\xi'}_C} |\tilde{\xi'}_C|^2+ m^2_{{\tilde{\xi}'}} |{\tilde{\xi}'}|^2 \,,
\end{eqnarray}
contains explicitly mass terms for $\tilde{\xi'}_C$ and  $\tilde{\xi'}$. At the same time the term $A_S (\phi_S\phi_S)'\chi$ breaks  the lepton number, the accidental  $SO(3)$, since $\chi$ is not a singlet of $SO(3)$, and also  the accidental $O(2)$ symmetry involving $\chi$ and $\theta$. If we assume the positivity of all the  vevs, we   ensure the stability of the potential  and  the positivity of the pseudo-Goldstone boson masses when  the soft terms   $m^2_T,\tilde{m}^2_T,A_T$ are negative while the soft terms  $A_S,m^2_{\tilde{\xi}'_C},m^2_{\tilde{\xi}'}$   positive and holds the condition
\begin{eqnarray}
(m_T^2+  \tilde{m}^2_T+2 A_T u_T)u_T^2+ A_S u_S^2 u_\chi +  m^2_{\tilde{\xi}'_C} u_{\tilde{\xi}'_C}^2  +  m^2_{\tilde{\xi}'} u_{\tilde{\xi}'}^2&<&0\,.
\end{eqnarray} 
In this way 
we guarantee the existence of a large  region in the parameter space for which the configuration of Eq.~\ref{eq:susycons} is the minimum of the potential of Eq.~\ref{eq:fullpot} and avoid the presence of massless particles. Notice that the soft term
$$ A_S [  (\phi_S\phi_S)'\chi+ H.c. ] $$
breaks also the discrete $Z_3$ symmetry  a part from the continuous leptonic number $U(1)_L$.  We can not care for its effects since the scale of $V_{soft}$ is many order of magnitude smaller than the scale of  the potential of Eq.~\ref{eq:susycons}.

\section{Corrections to the mass matrix textures}

In the following we will discuss the corrections to the leading mass matrix textures of Sec.~\ref{sec:model}. 
The corrections to the leading textures of Sec.~\ref{sec:model}  are induced by higher order operators in the superpotential of Eq.~\ref{eq:superP}  and are of two kinds:
{\begin{itemize}
\item[{\bf C.1}] that change the vacuum configuration given in Eq.~\ref{eq:susycons} and in particular disalign the triplets vevs; that directly affect the mass matrices, 
see sub-section\,\ref{sub1};
\item[{\bf C.2}] that directly affect the mass matrices, see sub-section\,\ref{sub2}.
\end{itemize} }

\subsection{Effects  in the flavon superpotential}\label{sub1}


In this section we address the problem of   the inclusion of the next to leading order operators in the superpotential of Eq.~\ref{eq:superP}.
We consider only the next to leading order operators  that  respect all the flavor 
symmetry $A_4\times Z_3\times Z_4$ and the continuous lepton number $U(1)_L$   since all the explicit breaking terms appear in $V_{soft}$ at a scale which is many order of magnitude lower than the flavor breaking scale.

To the full  leading  order  superpotential $W$
given in of  Eq.~\ref{eq:superP},  we add the next to  leading order part given by
\begin{equation}
W_{NL}=W_{{NL}_T}+W_{{NL}_S}+W_{{NL}_{S'}}\,,
\end{equation}
where the different terms read as
\begin{eqnarray}
\label{eq:potNL}
W_{{NL}_T} &=& \lambda_{T_1} \frac{1}{\Lambda}( \hat{\phi}_R^T \hat{\phi}_T)( \hat{\phi}_T\hat{\phi}_T)+ \lambda_{T_2} \frac{1}{\Lambda}( \hat{\phi}_R^T \hat{\phi}_T)'( \hat{\phi}_T\hat{\phi}_T)''+ \lambda_{T_3} \frac{1}{\Lambda}( \hat{\phi}_R^T \hat{\phi}_T)''( \hat{\phi}_T\hat{\phi}_T)' ,\nn\\
&+& \lambda_{T_4} \frac{1}{\Lambda}( \hat{\phi}_R^T \hat{\phi}_T  \hat{\phi}_T)_3\hat{\phi}_T ,\nn\\
W_{{NL}_S} &=& \lambda_{S_1} \frac{1}{\Lambda}( \hat{\phi}_R^S  \hat{\phi}_S \hat{\phi}_T)' \hat{\theta}+ \lambda_{S_2} \frac{1}{\Lambda}( \hat{\phi}_R^S  \hat{\phi}_S \hat{\phi}_T)''\hat{\chi}+\lambda_{S_3} \frac{1}{\Lambda}( \hat{\phi}_R^S \hat{\phi}_T)''\hat{\xi} \hat{\chi} + \lambda_{S_4} \frac{1}{\Lambda}( \hat{\phi}_R^S \hat{\phi}_T)'\hat{\xi} \hat{\theta} \nn\\
&+& \lambda'_{S_3} \frac{1}{\Lambda}( \hat{\phi}_R^S \hat{\phi}_T)''\hat{\xi'} \hat{\chi} + \lambda'_{S_4} \frac{1}{\Lambda}( \hat{\phi}_R^S \hat{\phi}_T)'\hat{\xi'} \hat{\theta}\,,\nn\\
W_{{NL}_{S'}} &=&\lambda_{\xi_1}\frac{1}{\Lambda} (\hat{\phi}^{'S}_R  \hat{\phi}_S  \hat{\phi}_S \hat{\phi}_T) +\lambda_{\xi_2} \frac{1}{\Lambda}(\hat{\phi}^{'S}_R \hat{\phi}_S \hat{\phi}_T)\hat{\xi} +\lambda'_{\xi_2} \frac{1}{\Lambda}(\hat{\phi}^{'S}_R \hat{\phi}_S \hat{\phi}_T)\hat{\xi'}\,.
\end{eqnarray}
The new superpotential is given by
\begin{equation}
W'= W+W_{NL}
\end{equation}
and the scalar potential with $V_{soft}=0$ is now
\begin{equation}
V=\left|\frac{\partial W'}{\partial f_{\phi_i}}\right|^2\,.
\end{equation}
The effect of $W_{NL}$ on the SUSY-conserving vacuum configuration of  Eq.~\ref{eq:susycons}  is just a shift in the vevs of the scalar fields and therefore the new vacuum
 configuration is given by
\begin{eqnarray}
\vev{\phi_T} &=&(v_T+\delta v_{T_1},v_T+\delta v_{T_2},v_T+\delta v_{T_3}),\nn\\
\vev{\phi_S} &=&(v_S+\delta v_{S_1},\delta v_{S_2},\delta v_{S_3}),\nn\\
\vev{\varphi_i}&=& u_i + \delta u_i\,,
\end{eqnarray}
where with  $\varphi_i$ we have indicated the generic $A_4$ singlet and with $u_i, \delta u_i$ its leading order vev and the shift  of the vev respectively. Since we expect the shifts to be subleading and scale as $\mathcal{O}(1/\Lambda)$, given the conditions 
\begin{equation}
\frac{\partial W'}{\partial f_{{\phi_i}}}_{|v_i+\delta v_i}=0\,,
\end{equation}
obtained by imposing a 
SUSY conserving vacuum, we expand them  linearly in the shifts
\begin{equation}
\label{eq:condNL} 
\frac{\partial W}{\partial f_{{\phi_i}}}_{|v_i+\delta v_i}+ \frac{\partial W_{NL}}{\partial f_{\phi_i}}_{|v_i}=0\,,
\end{equation}
and determine the shifts by solving the system given by Eq.~\ref{eq:condNL} .
For simplicity we give only the approximate solutions  obtained in the limit in which all  the couplings $\lambda_i$  are equal. The derivatives with respect to the components  $f_{\phi^T_{R_i}}$ give the following conditions
\begin{eqnarray}
\label{eq:delvev1}
\delta v_{T_1} &=\delta v_{T_2} &=\delta v_{T_3}= \delta v_{T}, \nn\\ 
\delta v_{T} &\sim& \frac{v_T^2}{\Lambda}\,.
\end{eqnarray}
The derivatives with respect to the components  $f_{\phi^S_{R_i}} $ give
\begin{eqnarray}
\label{eq:delvev2}
\delta v_{S_2}&=&\delta v_{S_3},\nn\\
\delta v_{S_1} &\sim& \frac{u_\chi u_\xi}{\Lambda},\nn\\
\delta v_{S_2} &\sim & \frac{v_Tu_\xi}{\Lambda}\,.
\end{eqnarray}
Finally the last derivatives give
\begin{eqnarray}
\label{eq:delvev3}
\delta u_{\xi} \sim \delta u_{\xi'} &\sim&  \frac{v_S v_T+ v_T u_\xi}{\Lambda}\,.
\end{eqnarray}
For the singlet $\xi,\xi'$ of $A_4$ we have to ensure that $\delta u_\xi/\Lambda\leq u_\xi/\Lambda$ ($\delta u_{\xi'}/\Lambda\leq u_{\xi'}/\Lambda$) and it is 
straightforward to check. For the triplets we need that the effects of the vev disalignments  due to the presence of 
the shifts do not destroy the TB predictions, or in other words, the corrections induced in 
$\sin \theta_{23},\sin\theta_{12},\sin \theta_{13}$ still keep in the interval indicated 
by the experimental data at the $3-\sigma$ level . 
As shown in \cite{Altarelli:2005yx} this constrains 
impose shifts that disalign the vevs to be of order 
\begin{equation}\label{eq:shift}
\frac{\delta v}{\Lambda}\leq \lambda^2 v\,,
\end{equation}
with $v$ the vev of the generic triplet.
In our case the shift of $\vev{\phi_T}$ maintains the correct alignment, so we only have to 
worry about   the disalignment of  $\vev{\phi_S}$. Since at the leading order its vev alignment is given by $(1,0,0)$, then we have only to consider 
 $\delta v_{S_2}$ and  $\delta v_{S_3}$.   Eqs.~\ref{eq:delvev1}--\ref{eq:delvev2}  and the assumptions done in 
the previous sections lead to
\begin{eqnarray}
\frac{\delta v_{S_{2,3}}}{\Lambda}&\sim&\lambda^3\,,
\end{eqnarray}
that satisfy the conditions in Eq.\,\ref{eq:shift}.

\subsection{Effects  in the Yukawa superpotential}\label{sub2}

Giving a look to the mass matrices of Eq.~\ref{eq:ml}--\ref{eq:mnu}--\ref{eq:mu} we see that highest operators giving contribute 
to mass matrices have scale as power of $1/\Lambda^3$. We have neglected operators of order greater than $1/\Lambda^3$.
To be consistent we should consider if the inclusion of  higher order operators 
may destroy the fermion mass hierarchies and the quark and lepton mixing matrices. 
For what concerns  the quark sector  the answer is trivial. 
For the up  quark the general entry may be written as
\begin{eqnarray}
\label{eq;genquark}
W_{q}&=&\hat{Q}_{a} \hat{U}^c_b \hat{H}_u \, \left( \lambda_{ab}\, \left(\frac{\hat{\chi}}{\Lambda}\right)^n  \left( \frac{\hat{\theta}}{\Lambda}\right)^m+ \lambda'_{ab}\, \left(\frac{\hat{\theta}}{\Lambda}\right)^r  \left( \frac{\hat{\xi}_C^2}{\Lambda^2}\right)^s \right)\,.
\end{eqnarray}
Here $n,m,r$ and $s$ are integers and $a,b$ are flavor indexes. We can obtain a  similar structure for the down quark.
Giving a look to the field content of Table~\ref{tab:scal} we see that the conservation of the lepton number in the matter sector imposes that  the first higher order operators are obtained by substituting $\hat{\chi}$ in Eq.~\ref{eq;genquark} with $(\hat{\phi}_T \hat{\phi}_T)'' \hat{\theta}/\Lambda^2$ or $\hat{\theta}$ with $(\hat{\phi}_T \hat{\phi}_T)' \hat{\chi}/\Lambda^2$. Since $u_\chi \sim \lambda u_\theta$ the latter is automatically suppressed. However also the first operator gives a very suppressed correction since  it goes as $ v_T^2 u_\theta/\Lambda^2 \sim \lambda^4 u_\chi $. 

For the  charged leptons the general Yukawa superpotential may be written as 
\begin{equation}
W_{l}= y_{a} \hat{E}^{c^a}_L\frac{(\hat{\phi}_T \hat{L}_L)^{r_a}}{\Lambda}  \hat{H}_d \left(\frac{\hat{\xi}_C}{\Lambda}\right)^n\,,
\end{equation}
that represents the compact expression of Eq.~\ref{eq:Ll}. Here $n$ is an integer, $a$ is the flavor index and $r_a$ indicates the different $A_4$ contractions 
of the triplets $\hat{\phi}_T$ and $\hat{L}_L$ that combine with the different $A_4$ singlets $\hat{E}^{c^a}$ to give an invariant. Due to the conservation of the $Z_4$ symmetry at order $1/\Lambda^{2+n}$ there is only one operator given by
\begin{equation}
\delta W_{l}= y_{a} \hat{E}^{c^a}_L\frac{(\hat{\phi}_T\hat{\phi}_T \hat{L}_L)^{r_a}}{\Lambda^2}  \hat{H}_d \left(\frac{\hat{\xi}_C}{\Lambda}\right)^n\,.
\end{equation}
This operator is potentially dangerous because changes the structure of the charged lepton mass matrix and as consequence the charged lepton mass matrix is no longer diagonalized by the $U$ given in Eq.~\ref{eq:ml} but by a $U'=U+\delta U$.  
This implies that  the lepton mixing matrix is deviated by the TB by an amount of order $\mathcal{O}(\delta U)$.  
However, since $\delta U \sim \mathcal{O}(v_T^2 /\Lambda^2)$, we obtain the relation 
$\delta U/U \sim \mathcal{O}(v_T /\Lambda)\sim\mathcal{O}(\lambda^2\div \lambda^3)$.
A deviation from the TB of this order is still compatible with the experimental data as indicated in the previous section.

For the neutrinos the general entry is given by
\begin{equation}
W_{l}= y_{\nu \xi}\frac{(\hat{L}  \hat{L})\, \hat{H}_u\hat{H}_u\hat{\xi}}{\Lambda^2}+
y_{\nu S}\frac{(\hat{\phi}_S\,\hat{L}\hat{L})\hat{H}_u \hat{H}_u}{\Lambda^2}+H.c. 
\end{equation}
and again the lepton number conservation implies that the only term that can give some dangerous corrections is given by
\begin{equation}
\label{eq:nextneut}
\delta W_{l}=\frac{\hat{\xi}\,\hat{ \phi_T}  (\hat{L}  \hat{L})\, \hat{H}_u\hat{H}_u}{\Lambda^3}.
\end{equation}
The same arguments used for the charged lepton mass matrix are applied here since the deviation  from the TB induced by Eq.~\ref{eq:nextneut} is again of order $ \mathcal{O}(v_T /\Lambda)\sim\mathcal{O}(\lambda^2\div \lambda^3)$, being $u_\xi\sim v_S$.

\end{document}